\documentclass[aps,prb,twocolumn,epsfig,superscriptaddress,showpacs,floatfix]{revtex4}
\usepackage[dvips]{epsfig}
\usepackage[dvips]{graphics}
\usepackage{bm}
\usepackage{times,color,amsmath,mathptmx}

\begin{document}
\hyphenation{na-no-tubes}

\title{Radial breathing mode of single-walled carbon nanotubes:  Optical transition energies and chiral-index assignment}

\author{J. Maultzsch}\email{janina@physik.tu-berlin.de}
\affiliation{Institut f\"ur Festk\"orperphysik, Technische Universit\"at Berlin, Hardenbergstr. 36, 10623 Berlin, Germany}
\author{H. Telg}
\affiliation{Institut f\"ur Festk\"orperphysik, Technische Universit\"at Berlin, Hardenbergstr. 36, 10623 Berlin, Germany}
\author{S. Reich}\thanks{permanent address: Department of Materials Science and Engineering, Massachusetts Institute of Technology, Cambridge, 02139 MA, USA}
\affiliation{Department of Engineering, University of Cambridge, Cambridge CB2 1PZ, United Kingdom}
\author{C. Thomsen}
\affiliation{Institut f\"ur Festk\"orperphysik, Technische Universit\"at Berlin, Hardenbergstr. 36, 10623 Berlin, Germany}
\begin{abstract}
We present a comprehensive study of the chiral-index assignment  of carbon nanotubes in aqueous suspensions by resonant Raman scattering of the radial breathing mode. We determine the   energies of the first optical transition in metallic tubes and of the second optical transition in semiconducting tubes for more than 50 chiral indices. The assignment is unique and does not depend on empirical parameters. The systematics of the so-called branches in the Kataura plot are discussed; many properties of the tubes are similar for members of the same branch. We show how the radial breathing modes observed in a single Raman spectrum can  be easily assigned based on  these systematics. In addition, empirical fits provide the energies and radial breathing modes for all metallic and semiconducting nanotubes with diameters between $0.6$ and $1.5$\,nm.  We discuss the relation between the  frequency of the radial breathing mode and  tube diameter. Finally, from the Raman intensities we obtain information on the electron-phonon coupling.
\end{abstract}
\maketitle

\section{Introduction}

Single-walled carbon nanotubes are tiny cylinders made out of carbon.\cite{reich04,iijima91,iijima93} They have many unique, fascinating properties.\cite{reich04} They are very strong and of extremely light weight, they are excellent conductors of heat and transport electrons ballistically. The properties of carbon nanotubes depend strongly on their microscopic structure, which is defined by the chiral index $(n_1,n_2)$. The best known example for this is that $2/3$ of all possible nanotubes are semiconductors and $1/3$ are metals.\cite{reich04}

Many applications of carbon nanotubes need one particular type of tube,  \emph{e.g.}, semiconducting tubes for transistors. Also, in fundamental studies we want to know which nanotube is probed experimentally. The growth of carbon nanotubes with a predefined microscopic structure remains a major challenge.  Therefore, the experimental determination of the chiral index is a current focus of carbon nanotube research.\cite{bachilo02,lebedkin03,miyauchi04,telg04,fantini04,jorio05} 
In principle, the chiral index of an individual tube can be determined by direct imaging techniques like scanning tunneling microscopy. However, the experimental error in the measurement of  diameter and chiral angle leads to uncertainties in the assignment of the chiral index. For example, the (13,1) tube with diameter $d=1.06$\,nm and chiral angle $\theta=3.7^\circ$ is geometrically close to the (14,1) tube [$d=1.14$\,nm and $\theta=3.4^\circ$], but the (13,1) tube is metallic and the (14,1) tube is semiconducting. 

Optical spectroscopy like  photoluminescence and Raman scattering uses properties that are different for each chiral index, \emph{i.e.}, electronic energies and phonon frequencies, to assign $n_1$ and $n_2$.
These methods are suitable also for macroscopic amounts of nanotubes.
At least two pieces of information are needed for an assignment, \emph{e.g.}, the combination of optical absorption and emission energies in photoluminescence and the phonon frequency plus one optical transition energy in Raman scattering. The assignment is then based on pattern recognition between experimental and theoretical data.~\cite{bachilo02,telg04,fantini04} Because of the systematics in the data of many nanotubes used for pattern recognition,  the assignment is stable against variations from small experimental errors.  

Optical methods are non-destructive and carry a large amount of information besides the information needed for the assignment itself. This has been demonstrated over the past three years by absorption and emission spectroscopy, time-resolved optical spectroscopy, and two-photon absorption.\cite{bachilo02,telg04,fantini04,bachilo03,lebedkin03,miyauchi04,huang04,ostojic04,wang04,jorio05,reich05,wang05,maultzsch05cm}
Because of the photoluminescence-based assignment suggested by Bachilo~\emph{et al.},\cite{bachilo02} the tubes selected in the optical studies were known. In this way, the electronic states, optical selection rules, carrier dynamics etc. of semiconducting tubes were studied as a function of the tube diameter and chiral angle.

The  advantage of Raman scattering over photoluminescence is that it can identify both metallic and semiconducting nanotubes.\cite{strano03b,telg04,fantini04,doorn04,thomsen05lss} Also, in semiconducting nanotubes Raman spectroscopy can be performed in resonance with the second optical transition, which is in the visible energy range. Thus, no infrared-sensitive spectrometers and detectors are needed. Finally, the Raman signal is more robust with respect to the environment of the nanotube. Photoluminescence, for example, is quenched in nanotube bundles by the presence of metallic tubes,\cite{reich05} whereas the Raman signal is still present. Raman spectroscopy holds  promise of identifying nanotubes in different environments with standard equipment.

The Raman experiments for an assignment of $n_1$ and $n_2$ reported so far were very laborious and required tunable lasers over wide energy ranges.\cite{strano03b,telg04,fantini04,doorn04} Now   a straightforward procedure to perform an assignment using one or two Raman spectra is needed. Also, we need to know by how much the environment of the tube affects the nanotube phonons and optical transition energies, because these two features are essential for an assignment based on  Raman scattering. First studies on the environment related effects concentrated on bundled tubes versus isolated surfactant-coated nanotubes in solution.\cite{oconnell04,fantini04} The interaction between the tubes in a bundle was found to shift the optical transition energies to the red as predicted by \emph{ab-initio} calculations by our group.\cite{reich02} In both experimental studies sodium dodecyl sulfate (SDS) was used as the surfactant for the debundled nanotubes. It will be important to know whether small changes in the tubes environment, \emph{e.g.}, a different surfactant, affect the phonon frequencies and optical transition energies.

Here we present a full analysis of the $(n_1,n_2)$ assignment from resonant Raman scattering of the radial breathing mode (RBM). We discuss the systematics of the so-called branches in the Kataura plot, where the optical transition energies are given as a function of the RBM frequency. The experimental data are extended by empirical fits  to a larger range of tube diameters. We show that our assignment is unique without additional parameters. From this assignment,  the coefficients $c_1$ and $c_2$ for the relation between diameter and RBM frequency are determined. By analyzing the intensity of the Raman signal, we observe systematic dependences of the Raman cross section on the chiral angle and on the nanotube family index $\nu=\pm1$. Our  results confirm recent \emph{ab initio} calculations of the matrix elements of  the electron-phonon interaction for the RBM.\cite{machon05} 
Changing the nanotube environment by using a different surfactant leads to variations in the RBM intensity and to small shifts in the optical transition energies. In metallic tubes we also observed small shifts in the RBM frequencies. These changes do not affect our pattern-based Raman assignment, but they show that the RBM frequency \emph{alone}---discarding the information about the Raman excitation energy---will never be sufficient to identify the chirality of a tube.
Finally, we explain the procedure for using our experimental and empirical data to assign the RBM peaks observed in a single Raman spectrum.

This article is organized as follows: in Sect.~\ref{sec_rbm} we give a brief overview over the radial breathing mode and  previous attempts for $(n_1,n_2)$ assignment based merely on the RBM frequency. The experimental methods are presented in Sect.~\ref{sec_exp}. In Sect.~\ref{sec_resonance_profiles} we explain the concept of resonant Raman scattering to determine the optical transition energies. The $(n_1,n_2)$ assignment is discussed in Sect.~\ref{sec_assign}. The experimental data are compiled and extended to arbitrary nanotubes using  empirical functions  in Sect.~\ref{sec_empirical}. In Sect.~\ref{sec_rbm_diameter} we derive the constants  $c_1$ and $c_2$ of the relation between RBM frequency and inverse diameter and discuss deviations from a linear behavior. 
In Sect.~\ref{rbm_intensities} the RBM intensity is analyzed as a function of the chiral angle. In Sect.~\ref{sec_sds_sdbs} we discuss surfactant-induced changes of the RBM spectra and transition energies. Finally, we give an instruction on how to assign the RBM peaks  in a Raman spectrum to $(n_1,n_2)$ in Sect.~\ref{sec_how_to}.

\section{Radial breathing mode}\label{sec_rbm}

\begin{figure}
\epsfig{bb=-100 0 651 512,file=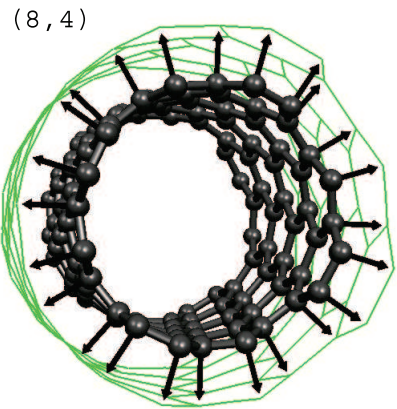,width=8cm,clip=}
\caption[]{Radial breathing mode of an (8,4) nanotube. The arrows show the phonon eigenvector. The RBM leads to a periodic increase and decrease of the tube diameter as shown by the wire model of the tube.}
\label{RBM_eigenvector}
\end{figure}
The radial breathing mode is the characteristic phonon mode of single-walled carbon nanotubes~\cite{reich04}. All atoms of the tube vibrate in-phase in the radial direction, see Fig.~\ref{RBM_eigenvector}.  A small non-radial component of the atomic displacement arises from mixing with the fully symmetric high-energy phonons.~\cite{dobardzic03,kuerti03,machon05} 
If the nanotube is approximated by a homogeneous cylinder, the frequency of the radial vibration is linear with the inverse tube diameter $1/d$ (Ref.~\onlinecite{reich04})
\begin{eqnarray}
\omega_{\mathrm{RBM}}=\frac{c_1}{d}+c_2\,.
\label{rbm_diameter}
\end{eqnarray}
The offset $c_2$ was originally introduced to account for additional external forces, \emph{e.g.}, from interactions with a substrate or neighboring tubes in a bundle.\cite{thomsen99c,venkateswaran99} On the other hand, changes in the environment of the tubes probed so far, lead to rather small changes in the RBM frequencies.\cite{oconnell04,fantini04} Therefore, $c_2$ should  be regarded more as a fitting parameter.
The geometrical diameter $d$ is given by 
\begin{eqnarray}
d=a_0\,\sqrt{n_1^2+n_1\,n_2+n_2^2}/\pi\, , 
\label{eq_diameter}
\end{eqnarray}
where $a_0=2.461\,${\AA} is the in-plane lattice constant of graphite. 
K\"urti \emph{et al.}\cite{kuerti03}  showed by  \emph{ab initio} calculations that deviations from Eq.~(\ref{rbm_diameter}) occur for small-diameter tubes, which additionally depend  on the chiral angle. These deviations are caused by first the small non-radial component of the vibration. Second,  the fully relaxed atomic structure of the tube has a slightly different diameter than the ideal geometrical diameter from Eq.~(\ref{eq_diameter}).\cite{kuerti03} 

The $\omega_{\mathrm{RBM}}$-diameter relation [Eq.~(\ref{rbm_diameter})] is often used to determine the tube diameters and the diameter distribution in a nanotube sample from Raman scattering. Comparing the diameter and Eq.~(\ref{eq_diameter}), the chiral indices are extracted,  giving an assignment of RBM frequencies to particular $(n_1,n_2)$ nanotubes.\cite{jorio01,bacsa02,kramberger03} This method requires the knowledge of the coefficients $c_1$ and $c_2$.  However, experimental and theoretical values of $c_1$ reported in the literature vary between 220 and 260\,cm$^{-1}\,$nm; $c_2$ varies between 0 and 20\,cm$^{-1}$, see for instance Refs.~\onlinecite{rao97,kuerti98,henrard99,jorio01}. For example, with $c_1=248$\,cm$^{-1}\,$nm and $c_2=0$\,cm$^{-1}$,  the observed RBM at 164\,cm$^{-1}$ was assigned to the (11,11) tube.\cite{jorio01} But if we choose a value obtained from \emph{ab initio} calculations  instead, $c_1=223$\,cm$^{-1}$, a different  peak (148\,cm$^{-1}$) is assigned to the (11,11) tube. 
Thus the assignment is very sensitive to the precise values of $c_1$ and $c_2$, and a particular RBM peak will be correlated with different $(n_1,n_2)$ depending on the variation of $c_1$ and $c_2$. 
An assignment based on $\omega_{\mathrm{RBM}}$ alone is in most cases not reliable, in particular for larger-diameter tubes with close-by RBM frequencies. It cannot be improved by higher accuracy in the experiment. Therefore, a second piece of information must be taken into account, as we show in Sect.~\ref{sec_assign}.

\section{Experimental methods}\label{sec_exp}
Raman experiments were performed on HiPCO-produced carbon nanotubes\cite{nikolaev99}, suspended in D$_2$O and wrapped by a surfactant (SDS, sodium dodecyl sulfate, and SDBS, sodium dodecylbenzene sulfonate).\cite{lebedkin03} The samples were excited by tunable Ti:sapphire and dye lasers and by an ArKr laser with powers of $\approx 15$\,mW focused into the nanotube solution. The scattered light was collected in backscattering geometry, dispersed by a Dilor XY800 triple monochromator and detected by a charge-coupled device.  The spectra were calibrated with a Neon lamp. We normalized the Raman intensity  with respect to the non-resonant Raman signal of CaF$_2$ and BaF$_2$ and to the laser power and integration time.

\section{Resonant Raman scattering}\label{sec_resonance_profiles}

In Raman scattering, the signal intensity increases strongly when the excitation energy approaches an allowed optical transition.\cite{cardona82,thomsen05lss} If the incoming or scattered light match the transition energy, this is called a resonance, and the intensity is at maximum. Recording the Raman intensity as a function of laser energy, we can determine the transition energies $E_{ii}$ in carbon nanotubes.  
The method is suitable for both semiconducting and metallic nanotubes, in contrast to photoluminescence, which probes only semiconducting tubes. By Raman spectroscopy we can directly probe the optical transition probability for $E_{ii}$, given the electron-phonon coupling is known. In contrast to photoluminescence, the strength of the signal is not additionally determined by the efficiency of absorption into other electronic bands and of  relaxation into dark and luminescent states.\cite{reich05b,jiang05}

\begin{figure} 
\begin{center}
\resizebox{8cm}{!}{
\includegraphics*{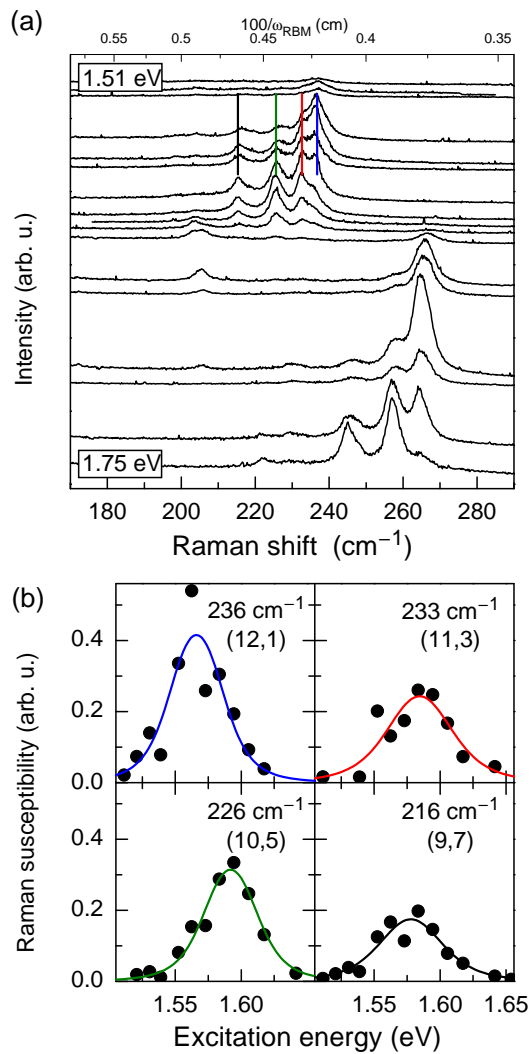}
}
 \end{center}
\caption{(Color online) (a) RBM spectra of carbon nanotubes at different excitation energies. The spectra are vertically offset for clarity. From top to bottom the laser energy increases between $1.51$\,eV and $1.75$\,eV. Each peak arises from a different $(n_1,n_2)$ nanotube. (b) Resonance profiles for the peaks marked in (a) by vertical lines. The dots are experimental data; the lines are fits according to Eq.~(\ref{resonance}).}
\label{laola}       
\end{figure}

In Fig.~\ref{laola}\,(a) we show the RBM spectra at different excitation energies. We see groups of several close-by peaks having their maximum strength one after the other, starting from the highest frequency and resembling a \emph{laola} wave~\cite{farkas02}. Each peak will be assigned to a different nanotube chirality $(n_1,n_2)$, see Sect.~\ref{sec_assign}. In Fig.~\ref{laola}\,(b) we show as an example the resonance profiles of four peaks belonging to the same group. The peak with the largest RBM frequency  has its resonance maximum at the lowest energy. The resonance energy increases as the RBM frequency decreases, and only for the last RBM peak, the resonance energy decreases slightly again. From the assignment (Sect.~\ref{sec_assign}) we find that such groups of RBM peaks form so-called branches in the Kataura plot. Each tube in a branch is related to its neighbor by $(n_1^\prime,n_2^\prime)=(n_1-1,n_2+2)$.\cite{telg04,thomsen05lss}

The Raman resonance profile is a superposition of an incoming and an outgoing resonance and can be described by\cite{cardona82}
%\begin{widetext}
\begin{multline}
I (E_l)=\Bigl(\frac{\mathcal{M}c}{\hbar\omega_{\mathrm{RBM}}}\Bigr)^2
\Big|\frac{1}{(E_l-E_{ii}-i\gamma/2)}-\\\frac{1}{(E_l-\hbar\omega_{\mathrm{RBM}}-E_{ii}-i\gamma/2)}\Big|^2\,,
\label{resonance}
\end{multline}
%\end{widetext}
where $E_l$ is the laser energy, $E_{ii}$ the energy of the allowed optical transition, and $\gamma$  the lifetime-broadening of the intermediate electronic states. $\mathcal{M}$ contains all matrix elements and $c$ summarizes all remaining factors. An incoming resonance occurs when $E_l=E_{ii}$, and an outgoing resonance when  $E_l=E_{ii}+\hbar\omega_{\mathrm{RBM}}$. If the incoming and outgoing resonances are not resolved in the resonance profile, the resonance maximum is at $\approx E_{ii}+0.5\,\hbar\omega_{\mathrm{RBM}}$.

Equation~\eqref{resonance} describes Raman scattering for a single resonant intermediate state $E_{ii}$. This corresponds to an excitonic transition, where the wave vector of the optically created exciton $Q=k_e+k_h$ is fixed by the momentum of the incoming photon $k_i=Q\approx 0$ ($k_e$ and $k_h$ are the wave vector of the electron and hole, respectively). Excitons have been shown to dominate optical transitions at room temperature in single-walled carbon nanotubes.\cite{wang05,maultzsch05cm} We, therefore, use resonant Raman scattering by excitons to describe our spectra.

We now briefly comment on the modifications of the resonant Raman cross section [Eq.~\eqref{resonance}] when considering band-to-band transitions, \emph{i.e.}, uncorrelated electrons and holes. A full discussion can be found in a review article by Thomsen and Reich\cite{thomsen05lss}. For band-to-band transitions, $E_{ii}$ is identified with the band gap of the resonant state instead of the $Q=0$ exciton energy. Additionally, in Eq.~\eqref{resonance} the square root of the incoming and outgoing resonance term has to be taken, see also Bussi~\emph{et al.}\cite{bussi05}. Despite these differences in the Raman matrix elements, the Raman profiles (squared matrix element) obtained for excitonic and band-to-band transitions are identical for all practical purposes. Only the lineshapes are slightly different, but they are indistinguishable in practice. Depending on the exact value of the electronic lifetime parameter, the experimental linewidth can also be different for band-to-band compared to excitonic resonances when using the same $\gamma$ in both calculations. Since $\gamma$ is not known independently, a Raman resonance profile cannot be used to discriminate between the two transition models, see Ref.~\onlinecite{thomsen05lss}. We stress that under no circumstances a resonant Raman profile with the lineshape of a square-root singularity (as first suggested in Ref.~\onlinecite{jorio01}) is expected.\cite{bussi05} 

Since excitons were found to dominate the optical spectra of semiconducting nanotubes by two-photon spectroscopy,\cite{wang05,maultzsch05cm} Eq.~\eqref{resonance} is certainly correct for semiconducting tubes. For metallic tubes, binding energies $\approx50\,$meV were predicted by first-principles calculations.\cite{spataru04} This binding energy is still more than twice the thermal energy at room temperature. It thus seems that metallic nanotubes also have excitonic resonances, because the Coulomb interaction is only screened along the nanotube axis. We therefore use Eq.~\eqref{resonance} to fit the resonance Raman profiles of both semiconducting and metallic tubes. We stress that the \emph{experimental} optical transition energies are not affected by the choice of the model.\cite{thomsen05lss} $E_{ii}$ are simply experimental values that need to be interpreted by a theoretical model.\cite{reich05b}

By fitting the resonance profiles  in Fig.~\ref{laola}\,(b) with Eq.~(\ref{resonance}), we obtain for each RBM peak the corresponding transition energy $E_{ii}$. 
For the broadening $\gamma$ we obtained  $\gamma\sim 0.06$\,eV for most transitions. The experimental data presented in Refs.~\onlinecite{fantini04} and \onlinecite{jorio05} from SDS-wrapped HiPCO tubes agree with ours to within experimental error.
In  Tables~\ref{tab_alles_sem} and \ref{tab_alles_met} we summarize all measured RBM frequencies and  optical transition energies $E_{ii}$.  

\section{$(n_1,n_2)$ assignment of RBM frequencies and transition energies}\label{sec_assign}

Having determined the pairs of RBM frequencies and transition energies, $(\omega_{\mathrm{RBM}}$, $E_{ii})$, we  show in this section how we assign them to particular tube chiralities $(n_1,n_2)$.\cite{telg04} The assignment is based on characteristic patterns in the experimental and theoretical data. We  do not require a \emph{quantitative} agreement between theory and experiment. Neither do we use  any calculated transition energies or RBM frequencies  nor the luminescence-based assignment suggested by Bachilo~\emph{et al.}\cite{bachilo02}.

\begin{figure} 
\begin{center}
\resizebox{8cm}{!}{
\includegraphics*{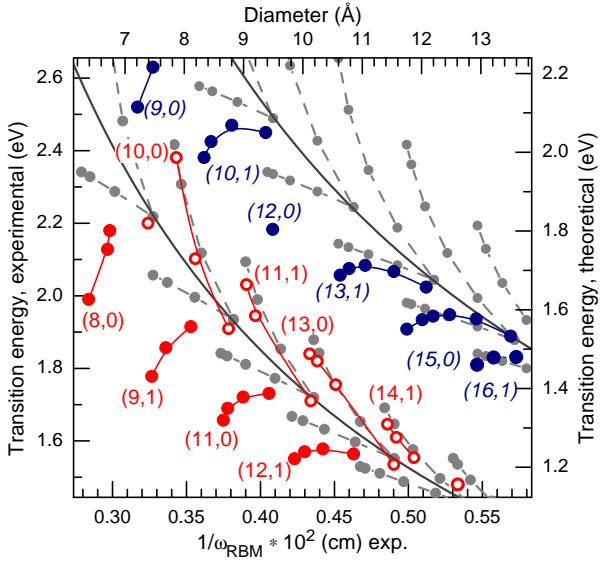}
}
 \end{center}
\caption{(Color online) Experimental (large colored circles, left and bottom axes) and theoretical (small gray circles, right and top axes) Kataura plot. The second transitions   of semiconducting tubes $E_{22}^S$  and the first transitions $E_{11}^M$ of metallic tubes   are shown.
The solid lines give the approximate $1/d$ dependence of the transition energies. 
 The dashed lines indicate the ``V''-shaped branches, where the chirality  of a tube is related to its left neighbor  $(n_1,n_2)$ by $(n_1^\prime,n_2^\prime)=(n_1-1,n_2+2)$.   In the experimental data, the assignment is given for the first tube in each branch, where upright (red) numbers indicate semiconducting and  italic (blue) numbers indicate metallic tubes. The semiconducting tubes are divided into two families with $\nu=(n_1-n_2)\,\textrm{mod }3=-1$ (full circles, lower branches) and with $\nu=+1$ (open circles, upper branches).}
\label{kataura_1}       
\end{figure}

The RBM frequency $\omega_{\mathrm{RBM}}$ is proportional to the inverse diameter, $\omega_{\mathrm{RBM}}\propto 1/d$ [Eq.~(\ref{rbm_diameter})]. We therefore obtain an experimental Kataura plot by plotting $E_{ii}$ as a function of  $1/\omega_{\mathrm{RBM}}$, see Fig.~\ref{kataura_1} (large red and blue circles).\cite{telg04,kataura99} We do not use any additional assumptions, in particular, the values of the coefficients $c_1$ and $c_2$ in Eq.~(\ref{rbm_diameter}) are unknown. 
The assignment is found by  comparing the experimental Kataura plot with a theoretical one (small gray circles in Fig.~\ref{kataura_1}). The theoretical transition energies are calculated from a third-nearest-neighbor tight-binding approximation, fit to density-functional-theory calculations (DFT).\cite{reich02b} To make both Kataura plots match we have to shift and stretch the axes of one Kataura plot with respect to the other. 

On the energy axis, the necessary shift and stretching reflects the uncertainties in the calculation of the optical transitions. The tight-binding calculation does not account for curvature effects, electron-electron, and electron-hole interaction. In particular, excitonic effects\cite{spataru04,kane04,perebeinos04,chang04,reich05b} have been shown to dominate the optical transitions with binding energies $\sim 400$\,meV.\cite{wang05,maultzsch05cm} Along the diameter axis, stretching of the Kataura plot corresponds to adjusting the unknown coefficient $c_1$ in Eq.~(\ref{rbm_diameter}); the shift  leads to the offset $c_2$. 

In the following we explain the procedure  and the criterion for a correct assignment in more  detail. 
The transition energies as a function of tube diameter follow roughly an $1/d$ dependence (solid lines in Fig.~\ref{kataura_1}). Chirality-dependent deviations from this behavior result in ``V''-shaped branches (dashed lines in Fig.~\ref{kataura_1}). The tubes with  largest chiral angle [armchair  $(n,n)$ or near-armchair $(n,n-1)$ direction] are at the inner position, \emph{i.e.}, closest to the $1/d$ line. At the outermost positions are the tubes with smallest chiral angle, \emph{i.e.}, zig-zag $(n,0)$ or near-zig-zag $(n,1)$ tubes. Starting there with chirality $(n_1,n_2)$, the neighboring tubes in the same branch (\emph{laola}) are given by 
\begin{eqnarray}
(n_1^\prime,n_2^\prime)=(n_1-1,n_2+2)
\end{eqnarray}
 with $n_1^\prime > n_2^\prime$. For example, the (12,1) \emph{laola} contains the following tubes: (12,1), (11,3), (10,5), (9,7), compare also Fig.~\ref{laola}.
The tubes belonging to the same branch can also be specified by $2\,n_1+n_2$ being constant.\cite{fantini04} 
Our assignment  makes use of these branches, requiring a good agreement between the patterns of the branches in theory and experiment. In particular, the number of tubes within a given  branch is unambiguously determined by the construction of a nanotube from a graphite sheet.\cite{reich04} 

Figure~\ref{kataura_1} shows the best match between our experimental and theoretical Kataura plot obtained by the above method. The assignment of our data points to the chiralities  $(n_1,n_2)$ directly follows from this plot and is indicated for the first tube of each branch. 
We assign all of our data to the second transition of semiconducting tubes, $E_{22}^S$, and to the first transition of metallic tubes, $E_{11}^M$.
For semiconducting tubes, we observe both the upper (open circles) and lower branches (full circles), whereas only the lower branches of metallic tubes are seen in the experiment.
Because of the systematics of the Kataura plot, tubes with similar diameter and chiral angle but from different branches, such as the metallic (13,1) and the semiconducting (14,1) tube, cannot be easily confused. 

The key point of the assignment by pattern recognition and pattern matching is that it greatly reduces the possible choices for an assignment. Take for instance the point at $1/\omega_\mathrm{RBM}=0.33\cdot10^{-2}$\,cm and $E_{22}=1.78\,$eV that is assigned to the (9,1) tube in Fig.~\ref{kataura_1}. We cannot assign this particular RBM to the (10,0) tube, because the (10,0) is at the wrong side of the $1/d$  line. We also see from the patterns that the (9,1) point has to be assigned to the outermost tube in a branch. 
This leaves us with two or three alternative assignments, where we  shift the entire experimental and theoretical plot with respect to each other. We  show in the next paragraph that such an attempt leads to contradictions between theory and experiment. We also discuss the possibility of assigning some of our data to different sets of transitions like $E_{11}^S$ instead of $E_{22}^S$.

\subsection{(Im)possible alternative assignments\label{sec_alternative}}

\begin{figure} 
\begin{center}
\resizebox{8cm}{!}{
\includegraphics*{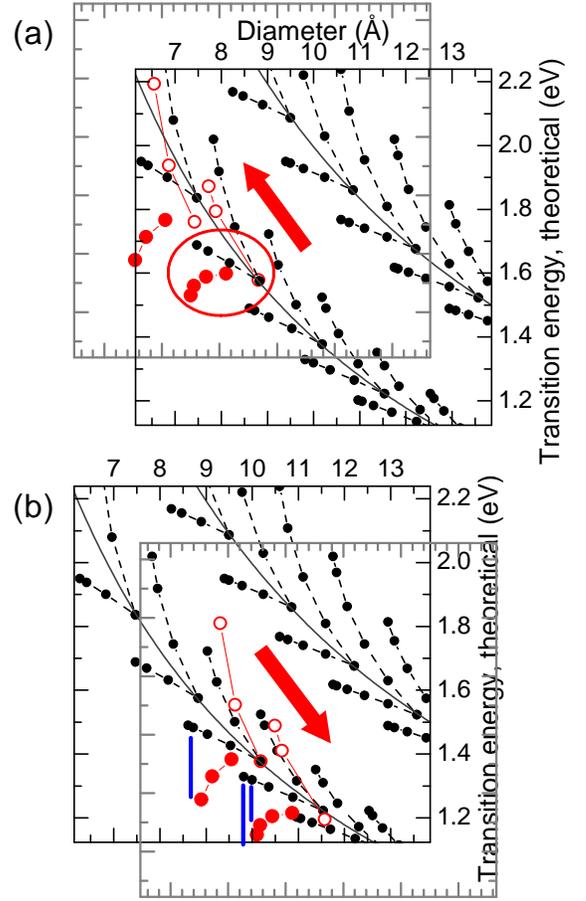}
}
 \end{center}
\caption{(Color online) Kataura plot showing two assignments that violate the pattern system and are therefore incorrect. (a) The experimental data were shifted ``up'' (to the upper left corner) when compared to Fig.~\ref{kataura_1}. Only two experimental branches are shown for clarity. The circle highlights the Kataura branch where this assignment produces five experimental RBMs for a branch that contains only four nanotubes. (b) same as (a), but for the shift down. The vertical lines highlight regions where the $1/d$ patterns of the RBM are  violated.}
\label{kataura_2}       
\end{figure} 

We first consider assigning our $(\omega_{\mathrm{RBM}},E_{ii})$ pairs to a different branch, keeping the overall assignment to the second transition in semiconducting and the first transition in metallic tubes.
Figure~\ref{kataura_2}\,(a) shows the combined experimental and theoretical Kataura plot when shifting the experimental data to the left or ``up'' by one branch. The upshift results in experimental branches where the number of RBMs is larger than the number of tubes in the branch [circle in Fig.~\ref{kataura_2}\,(a)]. This assignment is, therefore, incorrect. 

Figure~\ref{kataura_2}\,(b) is the plot for shifting the experimental data to the right or ``down''. 
Some branches then have less RBMs than tubes. This is quite possible if we have not detected all nanotubes in the diameter range. The assignment suggested in Fig.~\ref{kataura_2}\,(b), however, strongly  violates the patterns in the nanotube diameter distribution as we explain now. The diameter patterns are best seen  for the outermost members of each branch in Fig.~\ref{kataura_1}: because every second branch starts with a zig-zag tube, the distance between the two outermost points of the branches alternates between extremely close in $1/\omega_{\mathrm{RBM}}$ and slightly further apart when going from one branch to the next. This pattern is found in the experimental data as well, compare, for instance, the (9,1) and (12,1) branches in Fig.~\ref{kataura_1} (two outermost points more separated along the $x$ axis) to the (11,0) branch (two outermost points close together). The important argument here is again the \emph{pattern}, not the agreement or disagreement on an absolute $1/\omega_{\mathrm{RBM}}$ scale. 
We can, therefore, exclude the assignment of Fig.~\ref{kataura_2}\,(b) as well. A shift of the branches by more than one branch up or down increases the disagreement between theory and experiment. 

\begin{figure} 
\begin{center}
\resizebox{8cm}{!}{
\includegraphics*{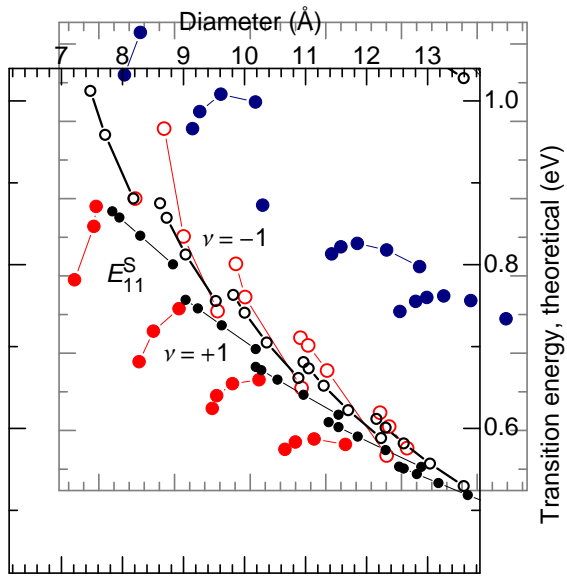}
}
 \end{center}
\caption{(Color online) Kataura plot showing a trial assignment where the experimental data are shifted down to the $E_{11}^S$ transitions. This assignment is incorrect for several reasons, see text. While  the experimental upper branches (open circles) match quite well the  pattern of the theoretical data, the lower branches disagree. The data assigned to the metallic tubes in Fig.~\ref{kataura_3} (dark blue circles) cannot be assigned to the second semiconducting transitions $E_{22}^S$, because they should be at exactly the same RBM frequencies (diameters) as the upper $E_{11}^S$ branches (open circles), which they are not.}
\label{kataura_3}       
\end{figure} 

One might also think of assigning the experimental data to different transitions $E_{ii}$, for example, shifting the experimental data down such that the measured RBM resonances correspond to the first transition of semiconducting tubes $E_{11}^S$ instead of the second transition $E_{22}^S$. The data assigned to metallic tubes in Fig.~\ref{kataura_1} then correspond either to $E_{22}^S$ or to $E_{11}^M$ (as before). Likewise, all data might be shifted up such that the $E_{22}^S$ data correspond to the metallic transitions $E_{11}^M$ and so forth. Although these alternative assignments are more of academic interest, since we know the $E_{22}^S$ transitions from photoluminescence,\cite{bachilo02} we want to discuss these ``exotic'' assignments briefly. We show that assigning our measured energies to other transitions than $E_{22}^S$ and $E_{11}^M$  systematically violates the Kataura-plot patterns and therefore result in incorrect assignments. It is an intriguing exercise highlighting very nicely the systematics in the Kataura plot and the pattern recognition idea.

Let us consider assigning the experimental $E_{22}^S$ data in Fig.~\ref{kataura_1} to the metallic $E_{11}^M$ resonance. The upper and lower branches of metallic transitions in the Kataura plot belong to the same chiralities. Thus the RBM frequencies in the upper and lower branches must be the same. This is in contrast with experiment, where the upper branches (open circles in Fig.~\ref{kataura_1}) begin and end at  larger diameter than the corresponding branches on the opposite side of the $1/d$ line. 

A downshift of the $E_{22}^S$ data to the $E_{11}^S$ transition is also impossible. When switching from a transition $E_{ii}^S$ to the next higher or lower transition, the family dependence is reversed. For $E_{22}^S$ the $+1$ tubes are above the $1/d$ line (see Fig.~\ref{kataura_1}); for $E_{11}^S$ the $\nu=+1$ tubes are below the $1/d$ line, \emph{vice versa} for the $\nu=-1$ tubes.\cite{reich04,reich00c} The two families start and end at different relative diameters for a given V-like curve. Changing the family dependence in the theoretical points by going from $E_{22}^S$ to $E_{11}^S$ completely disturbs the patterns.  In Fig.~\ref{kataura_3} we show such an attempt in detail. 
The former-assigned $E_{22}^S$ data are shifted down to the $E_{11}^S$ transitions. We stretched and displaced the Kataura plot to get the best match for the upper parts of the V-shaped curves, which now correspond to $\nu=-1$ tubes (open circles).  Obviously, the lower, $\nu=+1$, branches strongly violate the Kataura patterns. They are shifted to larger diameters with respect to the theoretical data. The situation gets even worse if we take the metallic tubes into account as well. Assigning them, \emph{e.g.}, to $E_{22}^S$ is impossible, because this results in nanotubes with the same chirality, but two different RBM frequencies.

We systematically considered other ``exotic'' assignments as well, \emph{e.g.}, the idea that certain data points in an experimental V-like curve are resonances coming from a different optical transition than the other points, say $E_{22}^S$ transitions are mixed with $E_{11}^S$ transitions in one and the same V-like curve in Fig.~\ref{kataura_1}. All these ideas can be excluded in a similar way as discussed for some selected alternative assignments in the preceeding paragraphs. 
The assignment in Fig.~\ref{kataura_1} is the only one that matches the systematics in the Kataura plot and obeys the $1/d$ patterns of the RBM frequencies. We summarize all measured RBM frequencies together with the assignment in Tables~\ref{tab_alles_sem} and \ref{tab_alles_met}.

Our Raman based assignment agrees with Bachilo's~\emph{et al.}\cite{bachilo02} suggestions from photoluminescence. They used a similar idea of pattern matching to correlate the theoretical transition energies with experiment.\cite{bachilo02,reich02b} The luminescence-based assignment was ambiguous; it needed an additional anchor element. There are two reasons why the $(\omega_{\mathrm{RBM}},E_{ii})$ pairs found by Raman scattering restrict the possible assignment more strongly than the ($E_{11},E_{22})$ pairs from luminescence: First, Raman scattering detects more nanotubes than photoluminescence, in particular, the metallic tubes and the tubes with small chiral angles at the end of the Kataura branches (zigzag and close-to zigzag tubes). Second, $\omega_{\mathrm{RBM}}$ is to good approximation independent of the chiral angle, whereas the $E_{ii}$ depend on diameter and chiral angle. This reduces the degrees of freedom for the ($\omega_{\mathrm{RBM}},E_{ii})$ pattern matching when compared to the $(E_{11},E_{22})$ matching and makes Raman scattering less ambiguous than photoluminescence. 

In summary, we assigned experimental RBM frequencies (together with $E_{ii}$) to particular nanotubes indices $(n_1,n_2)$. Based on this assignment, we correlate the RBM frequencies with tube diameters in Sect.~\ref{sec_rbm_diameter}. Our assignment is independent of empirical parameters.  The measured optical transition energies are excitonic energies, as known from recent experiments.~\cite{wang05,maultzsch05cm} As the assignment procedure does not rely on absolute energies when comparing with the theoretical Kataura plot, the results are  not affected by the strength of exciton binding in carbon nanotubes. In fact, our assignment is based on two patterns: The systematics in the distributions of tube diameters with $n_1$ and $n_2$, which comes from the $\bm{c}=n_1\bm{a}_1+n_2\bm{a}_2$ construction of a tube, and the family behavior of the nanotube transition energies.\cite{reich04,reich00c} As long as these two very general concepts in nanotube physics remain valid, our Raman based assignment is unique.

\begin{table*}
\caption{Summary of all observed RBM frequencies and transition energies of semiconducting tubes and their assignment. The tubes are grouped according to the branches (\emph{laola}) in the Kataura plot.  $\omega_{\mathrm{RBM}}^{\mathrm{emp}}$ and $E_{ii}^{\mathrm{emp}}$ give the empirical values from Eq.~(\ref{rbm_diameter}) with $c_1=215$\,cm$^{-1}$\,nm, $c_2=18$\,cm$^{-1}$, and from Eq.~(\ref{eq_sem}), respectively.  The index $\nu= (n_1-n_2)\,\textrm{mod}\,3=\pm 1$ indicates the two types of semiconducting tubes ($\nu=-1$ corresponds to $\nu=2$, as sometimes used in the literature). The diameters are calculated with $a_0=2.461$\,\AA . The nanotube sample contained SDS as surfactant for excitation energies above 1.99\,eV; below SDBS was used as surfactant. The experimental error of the transition energies is between 3\,meV (3 digits) and 30\,meV (2 digits). In the last column we give the Raman susceptibilty in arbitrary units, obtained by normalizing the maximum RBM peak area to integration time, laser power, relative spectrometer response and $\omega^4$. The intensity changes due to   different surfactants are taken into account, see Sects.~\ref{rbm_intensities} and \ref{sec_sds_sdbs}. The asterisks indicate susceptibility values with large experimental errors due to weak signal or incomplete resonance profiles.}\label{tab_alles_sem} 
\begin{ruledtabular}
\begin{tabular}{ccccccccccc}
$\nu$& \emph{laola}& $n_1$ & $n_2$ & $d$  & $\omega_{\mathrm{RBM}}$ & $E_{ii}$ & $\omega_{\mathrm{RBM}}^{\mathrm{emp}}$ &$E_{ii}^{\mathrm{emp}}$ &$\theta$ & suscept.\\
	 &             &       &       & (\AA )&      (cm$^{-1}$)        & (eV)     &    (cm$^{-1}$)        & (eV)     &       &      \\ \hline 
  $-1$& (12,1)  &    12 &     1 &  9.82 &  236.4 & 1.551 &237&1.54 & 4.0      & 4.1  \\
      &         &    11 &     3 & 10.00 &  232.6 & 1.570 &233&1.55 & 11.7     & 2.0  \\
      &         &    10 &     5 & 10.36 &  226.1 & 1.578 &225&1.56 & 19.1     & 2.3  \\
      &         &     9 &     7 & 10.88 &  216.0 & 1.564 &215&1.55 & 25.9     & 1.3  \\  \hline
                                                                                    
 $-1$ &  (11,0) &    11 &     0 &  8.62 &  266.7 & 1.657 &267&1.68 & 0.0      & 1.7 \\
      &         &    10 &     2 &  8.72 &  264.6 & 1.690 &264&1.69 & 8.9      & 2.3  \\
      &         &     9 &     4 &  9.03 &  257.5 & 1.72  &256&1.71 & 17.5     & 2.5  \\
      &         &     8 &     6 &  9.53 &  246.4 & 1.73  &243&1.72 & 25.3     & 1.4   \\  \hline
                                                                                     
 $-1$ &  (9,1)  &     9 &     1 &  7.47 &  306.2 & 1.78  &305&1.83 & 5.2      & 9.1 \\
      &         &     8 &     3 &  7.72 &  297.5 & 1.857 &296&1.88 & 15.3     & 35.8 \\
      &         &     7 &     5 &  8.18 &  283.3 & 1.915 &281&1.92 & 24.5     & 18.3 \\  \hline                                                                             
 $-1$ & (8,0)   &     8 &     0 &  6.27 &  352.2 & 1.99  &361&1.97 & 0.0      & 0.1 \\
      &         &    7  &     2 &  6.41 &        &       &353&2.03 &12.2      &      \\
      &         &    6  &     4 &  6.83 &        &       &333&2.15 &23.4      &      \\  \hline
      
 $+1$ & (14,1)  &    14 &     1 & 11.38 &  205.8 & 1.646 &207&1.66 &  3.4     & 0.3        \\
      &         &    13 &     3 & 11.54 &  203.3 & 1.610 &204&1.62 &  10.2    & 0.6        \\
      &         &    12 &     5 & 11.85 &  198.5 & 1.554 &199&1.56 &  16.6    & 0.1         \\ 
      &         &    11 &     7 & 12.31 &        &       &192&1.47 &  22.7    &      \\
      &         &    10 &     9 & 12.90 &        &       &185&1.38 &  28.3    &      \\ \hline
                                                                                   
 $+1$ & (13,0)  &    13 &     0 & 10.18 &  230.8 & 1.84  &229&1.85 &  0.0     & 0.2$^*$       \\
      &         &    12 &     2 & 10.27 &  228.1 & 1.82  &227&1.82 &  7.6     & 12.2$^*$      \\
      &         &    11 &     4 & 10.54 &  221.8 & 1.76  &222&1.74 & 14.9     &  0.1$^*$      \\
      &         &    10 &     6 & 10.97 &        &       &214&1.64 & 21.8     &      \\
      &         &     9 &     8 & 11.54 &  204.0 & 1.535 &204&1.52 & 28.1     & 0.5      \\  \hline
      
 $+1$ & (11,1)  &    11 &     1 &  9.03 &  256.0 & 2.031 &256&2.06 &  4.3     & 9.8     \\
      &         &    10 &     3 &  9.24 &  252.1 & 1.945 &251&1.98 &  12.7    & 3.6      \\
      &         &    9  &     5 &  9.63 &        &       &241&1.84 &  20.6    &      \\
      &         &     8 &     7 & 10.18 &  230.4 & 1.710 &229&1.69 &  27.8    & 0.4      \\  \hline
                                                                                             
 $+1$ & (10,0)  &    10 &     0 &  7.83 &  291.4 & 2.38  &292&2.35 &   0.0    & -       \\
      &         &    9  &     2 &  7.95 &        &       &288&2.28 &   9.8    &      \\
      &         &     8 &     4 &  8.29 &  280.9 & 2.10  &277&2.11 &  19.1    & 0.3      \\
      &         &     7 &     6 &  8.83 &  264.2 & 1.909 &261&1.91 &  27.5    & 2.6      \\  \hline
      
 $+1$ & (8,1)   &     8 &     1 &  6.69 &        &       &339&2.69 &   5.8    &     \\ 
      &         &    7  &     3 &  6.96 &        &       &327&2.48 &  17.0    &      \\ 
      &         &     6 &     5 &  7.47 &  308.6 & 2.20  &305&2.18 &  27.0    &  -  
 \end{tabular}
\end{ruledtabular}
\end{table*}
 
\begin{table*}
\caption{Summary of all observed RBM frequencies and transition energies of metallic tubes  and their assignment. For metallic tubes, $\nu = 0$. Except for armchair tubes, each metallic tube has two close-by transition energies\cite{reich00c} of which always the lower one was observed in the experiment.  }\label{tab_alles_met} 
\begin{ruledtabular}
\begin{tabular}{ccccccccccc}
$\nu$& \emph{laola}& $n_1$ & $n_2$ & $d$  & $\omega_{\mathrm{RBM}}$ & $E_{ii}$ & $\omega_{\mathrm{RBM}}^{\mathrm{emp}}$ &$E_{ii}^{\mathrm{emp}}$ &$\theta$ & suscept.\\
	 &             &       &       & (\AA )&      (cm$^{-1}$)        & (eV)     &    (cm$^{-1}$)        & (eV)     &       &      \\ \hline   
 0    & (16,1)  &    16	&     1 & 12.94 &  182.0 & 1.81  &  184 & 1.81   &  3.0 & 0.5 \\
      &         &    15 &     3 & 13.08 & 179.0  & 1.83  & 182  & 1.81 & 8.9 & 1.8 \\
      &         &    14 &     5 & 13.36 &  174.5 & 1.83  &  179 & 1.81 &  14.7 & 1.9 \\\hline
  0   & (15,0)  &    15 &     0 & 11.75 &  200.4 & 1.908 &201&1.93 & 0.0      & 0.7   \\
      &         &    14 &     2 & 11.83 &  196.3 & 1.934 &200&1.94 & 6.6      & 3.6  \\
      &         &    13 &     4 & 12.06 &  193.5 & 1.944 &196&1.94 & 13.0     & 2.5 \\
      &         &    12 &     6 & 12.44 &  189.4 & 1.948 &191&1.93 & 19.1     & 1.9  \\
      &         &    11 &     8 & 12.94 &  183.2 & 1.936 &184&1.92 & 24.8     & 1.9  \\
      &         &    10 &    10 & 13.57 &  175.7 & 1.889 &176&1.88 & 30.0     & 0.9  \\  \hline
                                                                                     
   0  &  (13,1) &    13 &     1 & 10.60 &  220.3 & 2.057 &221&2.07 & 3.7      & 1.5  \\
      &         &    12 &     3 & 10.77 &  217.4 & 2.075 &217&2.07 & 10.9     & 2.6  \\
      &         &    11 &     5 & 11.11 &  212.4 & 2.084 &211&2.08 & 17.8     & 1.7  \\
      &         &    10 &     7 & 11.59 &  204.0 & 2.067 &203&2.07 & 24.2     & 0.9  \\
      &         &     9 &     9 & 12.21 &  195.3 & 2.02  &194&2.04 & 30.0     & 0.5   \\  \hline 
                                 
 0    & (12,0)  &    12 &     0 &  9.40 &  244.9 & 2.18  &247&2.21 &  0.0     &  -       \\
      &         &    11 &     2 &  9.50 &        &       &244&2.22 &  8.2     &      \\
      &         &    10 &     4 &  9.78 &        &       &238&2.24 & 16.1     &      \\ 
      &         &    9  &     6 & 10.24 &        &       &228&2.25 & 23.4     &      \\ 
      &         &    8  &     8 & 10.85 &        &       &216&2.23 & 30.0     &      \\ \hline
      
 0    & (10,1)  &    10 &     1 &  8.25 &  276.3 & 2.38  &278&2.35 &  4.7     &   -       \\
      &         &     9 &     3 &  8.47 &  272.7 & 2.43  &272&2.40 &  13.9    &   -      \\
      &         &     8 &     5 &  8.90 &  262.7 & 2.47  &259&2.45 &  22.4    &   -      \\
      &         &     7 &     7 &  9.50 &  247.8 & 2.45  &244&2.46 &  30.0    &   -      \\  \hline 
                     
  0   &  (9,0)  &     9 &     0 &  7.05 &        &       &323&2.47 &   0.0    &  \\ 
      &         &     8 &     2 &  7.18 &  315.5 & 2.52  &317&2.53 &  10.9    & -  \\
      &         &     7 &     4 &  7.55 &  305.4 & 2.63  &302&2.65 &  21.1    & -\\ 
      &         &     6 &     6 &  8.14 &        &       &282&2.71 &  30.0    &          
\end{tabular}
\end{ruledtabular}
\end{table*}

\section{Transition energies of semiconducting and metallic tubes}\label{sec_empirical}

\begin{figure} 
\begin{center}
\resizebox{8cm}{!}{
\includegraphics*{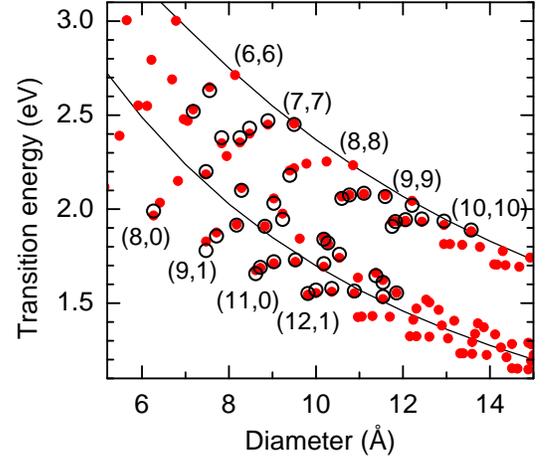}
}
 \end{center}
\caption{Experimental transition energies (open symbols) and empirical values from Eqs.~(\ref{eq_sem}) and (\ref{eq_met}) (closed symbols). For the metallic branches, the inner (armchair) tube is indicated. }
\label{empirical}       
\end{figure}

In this section we provide empirical fits to the experimental transition energies $E_{22}^S$ and $E_{11}^M$ in order to apply these expressions to chiralities not observed in the experiment.  
We start with the $1/d$ relation for the band gap, $E_{ii}=2i\gamma_0\,a_{\mathrm{C-C}}/d$,\cite{mintmire98,reich04} expanding it in $1/d$. We add a chiral-angle dependent term to model the branches of the Kataura plot, which is larger for smaller diameters.\cite{reich00c,saito00}
For the $E_{22}^S$ transitions of semiconducting tubes, 
\begin{eqnarray}
E_{22}^S=\gamma_0\,(\frac{4\,a_{\mathrm{C-C}}}{d}+\gamma_1\frac{a_{\mathrm{C-C}}^2}{d^2})+\nu\,\gamma_2\,\cos(3\theta)\frac{a_{\mathrm{C-C}}^2}{d^2},
\label{eq_sem}
\end{eqnarray}
with the parameters $\gamma_0$, $\gamma_1$, and $\gamma_2$. $a_{\mathrm{C-C}}$ is the length of the carbon-carbon bonds ($a_{\mathrm{C-C}}=a_0/\sqrt{3}$), and $\nu=(n_1-n_2)\,\textrm{mod}\,3$ is the family index taking the values $\pm1$ in semiconducting tubes and $\nu=0$ in metallic tubes. $\theta$ is the chiral angle of the $(n_1,n_2)$ tube, being zero in zig-zag tubes.\cite{reich04}
From the fit to our data we obtain $\gamma_0=3.53$\,eV, $\gamma_1=-4.32$, and $\gamma_2=8.81$\,eV. Analogously, we approximate the transition energies in metallic tubes by 
\begin{eqnarray}
E_{11}^M=\gamma_0\,(\frac{6\,a_{\mathrm{C-C}}}{d}+\gamma_1\frac{a_{\mathrm{C-C}}^2}{d^2})-\gamma_2\,\cos(3\theta)\frac{a_{\mathrm{C-C}}^2}{d^2},
\label{eq_met}
\end{eqnarray}
and find $\gamma_0=3.60$\,eV, $\gamma_1=-9.65$, and $\gamma_2=11.69$\,eV. The empirically determined  energies are shown as a function of diameter in Fig.~\ref{empirical}, where $d$ is given by $(n_1,n_2)$ [Eq.~(\ref{eq_diameter})]. Because the electronic properties change dramatically for small-diameter tubes\cite{machon02,liu02}, the empirical data given here are valid only for tubes with diameters $d\geq 6$\AA . Our empirical extrapolation fits well with the experimental $E_{22}^S$ data obtained by Doorn \emph{et al.}\cite{doorn04} for the tubes outside the range of our experiments.  All empirical transition energies are listed in the supplementary material, Ref.~\onlinecite{epaps}.

The $\gamma_0$ obtained from Eqs.~\eqref{eq_sem} and \eqref{eq_met} are quite close to the corresponding parameter in graphite (3.2\,eV, Ref.~\onlinecite{toy76}) and much larger than assumed for carbon nanotubes in the past ($2.5-2.75\,$eV, Refs.~\onlinecite{jorio01,mintmire93}). From our fits it seems that $\gamma_1$ scales with the square of $i$ ($E_{11}=E_{11}^S$, $E_{22}=E_{22}^S$, $E_{33}=E_{11}^M$,  see Refs.~\onlinecite{reich04,mintmire98}), whereas the chirality-dependent correction shows only a linear scaling. These corrections reflect a mixture of trigonal warping of the graphene band structure,\cite{reich00c} curvature effects,\cite{reich02,popov04b} and exciton effects\cite{kane04,perebeinos04}. Very recently, first experimental results on the exciton binding energies in carbon nanotubes were published.\cite{wang05,maultzsch05cm} Although Maultzsch~\emph{et al.}\cite{maultzsch05cm} reported a family dependence of the exciton binding energies and also a dependence on the size of the nanotube band gap, it is difficult to quantify these effects from the data at hand. Perebeinos~\emph{et al.}\cite{perebeinos04} suggested a scaling of the exciton binding energy with the effective mass of the electrons and the hole as $E_b\propto m^{\alpha-1}$, where $\alpha=1.4$ was found from tight-binding calculations. The effective mass in turn depends on the family of a tube and its diameter and chiral angle. Nevertheless, for the time being we prefer to understand Eqs.~\eqref{eq_sem} and \eqref{eq_met} as empirical functions instead of giving the parameters and dependences a strict physical meaning in terms of curvature and exciton effects. The two relations fit the experimental data very well, see Fig.~\ref{empirical}. It would be interesting to obtain Raman data for the first optical transition energy to see whether the proposed scaling for $\gamma_1$ and $\gamma_2$ applies to this transition as well.

Strano \emph{et al.}\cite{strano03b,strano03d} used a similar expression to fit the observed transition energies in semiconducting and metallic tubes. Our results for semiconducting tubes are in good agreement with the empirical values in Refs.~\onlinecite{strano03b} and \onlinecite{strano03d}. For metallic tubes, in contrast, we find large deviations between our experimental data and the empirical  data in Ref.~\onlinecite{strano03b}. In particular, the Raman resonances of some metallic tubes were assigned to the \emph{second} metallic transitions in Ref.~\onlinecite{strano03b}, whereas from our data it is obvious that only the \emph{first} metallic transitions are observed. For example, the resonance of the (7,7) armchair tube (see Fig.~\ref{empirical}), \emph{i.e.}, at the inner position of a ``V''-shaped branch, is in Ref.~\onlinecite{strano03b} assigned to the second resonance of the (12,0) zig-zag tube. Therefore, the empirical expressions given by Strano \emph{et al.}\cite{strano03b} underestimate the transition energies of metallic tubes.

In Ref.~\onlinecite{jorio05} a detailed comparison between the experimental transition energies with  tight-binding results can be found. The authors  present diameter and chirality-dependent corrections involving eight fitting parameters for each set of transitions. 

\section{Relation between $\omega_{\mathrm{RBM}}$ and diameter}\label{sec_rbm_diameter}

In contrast to previous attempts to obtain $(n_1,n_2)$ from the Raman spectrum of a tube\cite{jorio01,kramberger03} we first assigned an RBM frequency to a nanotube. We used the fact that the RBM frequency is approximately linear with the inverse tube diameter. We did, however, not include Eq.~(\ref{rbm_diameter}) explicitly, in particular, no values $c_1$ and $c_2$ were given. 
From this assignment we now calculate the tube diameter and fit $c_1$ and $c_2$ from the $\omega_\mathrm{RBM}$ \emph{versus} diameter plot. The key difference to other work is that a particular Raman line is always assigned to  the same nanotube within our approach. Using, \emph{e.g.}, the diameters reported by K\"urti~\emph{et al.}\cite{kuerti03} from first-principles calculations instead of the geometrical expression in Eq.~\eqref{eq_diameter} we find slightly different numbers for $c_1$ and $c_2$. The assignment in Tables~\ref{tab_alles_sem} and \ref{tab_alles_met}, however, remains the same. For an assignment it is usually better to work with the Tables instead of Eq.~\eqref{rbm_diameter} as we discuss in Sect.~\ref{sec_how_to}.

\begin{figure} 
\begin{center}
\resizebox{8cm}{!}{
\includegraphics*{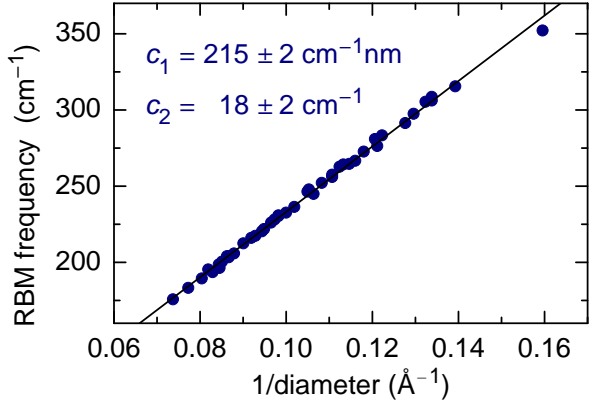}
}
 \end{center}
\caption{Linear fit of the observed RBM frequencies as a function of inverse tube diameter $1/d$. The diameter of the assigned nanotubes is calculated from Eq.~(\ref{eq_diameter}) with $a_0=2.461$\,\AA .}
\label{linear_fit}       
\end{figure}

In Fig.~\ref{linear_fit} we show a linear fit to our data points according to Eq.~(\ref{rbm_diameter}).
Using $a_0=2.461$\,\AA, we obtain $c_1=215$\,cm$^{-1}$nm and $c_2=18$\,cm$^{-1}$.
With $a_{\mathrm{C-C}}=a_0/\sqrt{3}=1.44$\,\AA, also used in the literature,  $c_1=218$\,cm$^{-1}$nm and $c_2=18$\,cm$^{-1}$. The coefficients are thus very sensitive to how the tube diameter is determined. Therefore, they provide an estimate of the RBM frequency for a given diameter [Eq.~(\ref{rbm_diameter})], but should not be used to compare (or even assign) an experimental RBM frequency to a nanotube diameter with high accuracy.

Deviations from the linear dependence of the RBM frequency on the inverse diameter have been predicted for small-diameter tubes by K\"urti \emph{et al.}\cite{kuerti03} from first-principles calculations. In Raman experiments, based on the assignment of Ref.~\onlinecite{bachilo02}, Jorio \emph{et al.}\cite{jorio05}  observed  deviations of a few wavenumbers, depending slightly on the chiral angle. 
Here we show explicitly that the behavior of the experimental RBM agrees very well with the calculations in Ref.~\onlinecite{kuerti03}.

\begin{figure} 
\begin{center}
\resizebox{8cm}{!}{
\includegraphics*{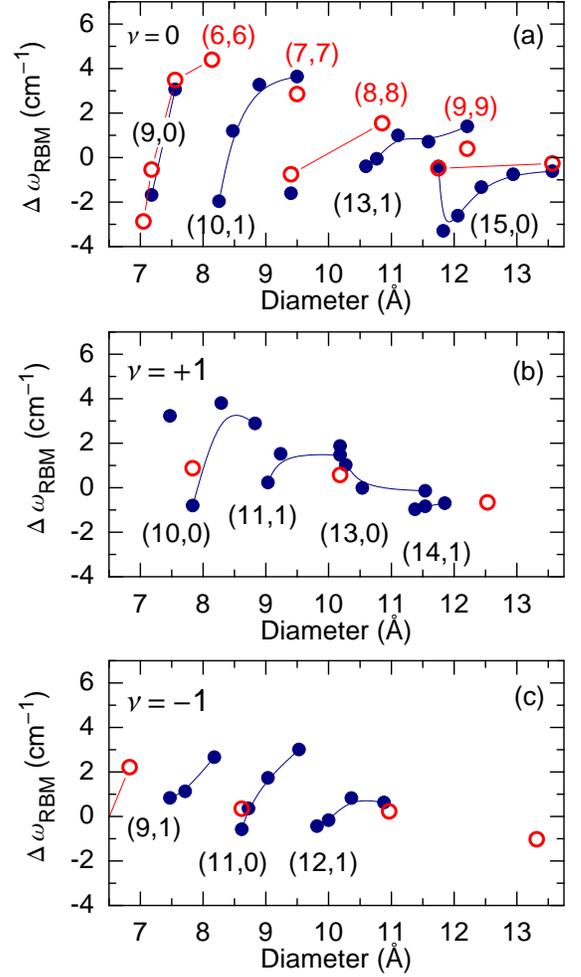}
}
 \end{center}
\caption{Difference between experimental RBM frequencies and calculated RBM frequencies from Eq.~(\ref{rbm_diameter}) with $c_1=215$\,cm$^{-1}$nm and $c_2=18$\,cm$^{-1}$ (full circles). The lines connect tubes of the same branch, labeled by the first member according to the Kataura plot and Tables~\ref{tab_alles_sem} and \ref{tab_alles_met}. (a), (b), and (c) show the tubes with $\nu=0$, $\nu=+1$, and $\nu=-1$, respectively. The chiral angle within each branch decreases with  decreasing diameter. Open (red) circles show the difference between $\omega_\mathrm{RBM}^\mathrm{DFT}$ from first-principles calculations\cite{kuerti03} and $\omega_\mathrm{RBM}$ from a linear fit to these theoretical data. In  (a) the armchair tubes are indicated.}
\label{rbm_abweichung}       
\end{figure}

Figure~\ref{rbm_abweichung} shows the difference $\Delta\,\omega_\mathrm{RBM}$ between the experimental $\omega_\mathrm{RBM}$ and the empirical values calculated from Eq.~(\ref{rbm_diameter}) with $c_1=215$\,cm$^{-1}$nm and $c_2=18$\,cm$^{-1}$ (full circles). In general, the deviations from the linear fit  increase for smaller diameters, in agreement with the predictions. They vary between $+4$ and $-2$\,cm$^{-1}$, depending on the chiral angle. For tubes with $d\leq 10$\,\AA , $\Delta\,\omega_\mathrm{RBM}$ has a large and positive value for (near-)armchair tubes and decreases to negative values for (near-) zig-zag tubes. 

K\"urti \emph{et al.}\cite{kuerti03} showed that isolated armchair tubes follow the linear relation with the smallest deviations, whereas zig-zag tubes have the largest (negative) deviation.  If we assume the line connecting the armchair tubes in Fig.~\ref{rbm_abweichung}\,(a) to be the $\Delta\,\omega_\mathrm{RBM}=0$ line,\cite{fussnote_rbm} we observe the same trend of increasing deviation towards zig-zag tubes. This agrees with the prediction that zig-zag tubes show the strongest rehybridization effects\cite{reich02} and the largest increase in bond length\cite{kuerti03}, both effects resulting in a weakening of the RBM frequency.

To compare with the theoretical data of Ref.~\onlinecite{kuerti03} quantitatively, we performed a linear fit to the \emph{ab initio} values $\omega_\mathrm{RBM}^\mathrm{DFT}$ and analyzed the differences between $\omega_\mathrm{RBM}^\mathrm{DFT}$ and this linear fit (open circles in Fig.~\ref{rbm_abweichung}). We find a very good quantitative agreement between experiment and theory, although the calculations were performed for isolated tubes in vacuum. This confirms that the deviations from the linear relation are mostly due to changes in the strength of the bonds (rehybridization and bond lengths). 

\section{RBM intensities}\label{rbm_intensities}
The Raman resonance profile given by Eq.~(\ref{resonance}) accounts for the position of the resonance maximum, which we discussed so far. In this section, we evaluate the relative  strength of the Raman signal to obtain information on the matrix elements $\mathcal{M}$. The  matrix elements $\mathcal M $ consist of the electron-photon coupling, $\mathcal M_{e-r}$, and the electron-phonon coupling, $\mathcal M_{e-ph}$, and  $|\mathcal M|^2=|\mathcal M_{e-r}\,\mathcal M_{e-ph}\,\mathcal M_{e-r}|^2$. The constant $c$ contains the remaining factors such as response of the spectrometer and $\omega^4$ dependence of the Raman cross section (taken into account by normalization to the non-resonant Raman signal of BaF$_2$ and CaF$_2$), laser power, integration time, scattering volume, and concentration of the nanotube solution. We do not make any assumptions on the diameter and chirality distribution in our sample, thus showing the bare Raman intensities. The diameter distribution can be determined by electron miscroscopy\cite{arcos05} or electron diffraction\cite{liu05}. As we show below, the relative abundance of particular chiralities cannot be determined by the bare Raman or luminescence intensities.

\begin{figure} 
\begin{center}
\resizebox{8cm}{!}{
\includegraphics*{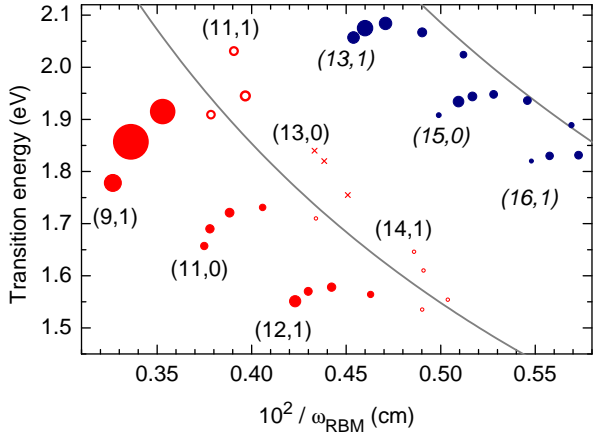}
}
 \end{center}
\caption{(Color online) Part of the experimental Kataura plot, where the intensity of the Raman signal is given by the area of the circles. Grey (red) and dark (blue) symbols indicate semiconducting and metallic tubes, respectively. The smallest circles account for all values below 0.5 in Table~\ref{tab_alles_sem}. The crosses indicate tubes with large uncertainty in the Raman intensity, see also Table~\ref{tab_alles_sem}.  In addition to the normalization procedure described in the text, the intensity of the (11,1) tube is divided by 3.4 and the intensity of the (13,1) branch is multiplied by three. This accounts for the changes in the Raman signal strength due to a different surfactant for the sample used above $E_l=1.99$\,eV, see Sect.~\ref{sec_sds_sdbs}.}
\label{kandinsky}       
\end{figure}

In Fig.~\ref{kandinsky} we show part of the experimental Kataura plot, where the area of the circles indicates the strength of the Raman signal. We observe two  trends: first, the Raman cross section increases for smaller diameter; second, it is in general much larger for the lower branches ($\nu=-1$ for $E_{22}^S$) than for the upper branches. The upper branches  of metallic tubes were not observed at all. Both the diameter dependence and the $\nu=\pm1$ dependence  were predicted by first-principles calculations of the electron-phonon coupling matrix elements $\mathcal M_{e-ph}$,\cite{machon05} see also Ref.~\onlinecite{popov04}. 

The electron-phonon coupling for the RBM becomes stronger for smaller-diameter tubes, because the same radial displacement results in a larger change of the carbon-carbon bonds in smaller tubes. The dependence on $\nu=\pm1$ can be understood within the zone-folding picture. The index $\nu=\pm1$ indicates from which  side of the $K$ point in the graphene Brillouin zone the electronic states are derived. In graphene, the coupling between electrons close to the $K$ point and phonons analogous to the RBM is larger between the $K$ and $M$ points than in the $K$-$\Gamma$ direction.\cite{machon05} 
Similar results from Raman scattering  were reported  by Doorn \emph{et al.}~\cite{doorn04} and Jorio  \emph{et al.}~\cite{jorio05};  calculations within an empirical tight-binding description give the same dependence on $\nu=\pm1$.\cite{popov04,goupalov05}

According to the \emph{ab-initio} calculation by Mach\'on~\emph{et al.},\cite{machon05} we expect the opposite family behavior for Raman scattering in resonance with the \emph{first} transition energy of  semiconducting tubes. For this resonance, the Raman susceptibility of the $\nu=+1$ nanotubes should be larger than the susceptibility of the $\nu=-1$ nanotubes. Once again, this can be understood in the zone-folding approach, because the first and second optical transition in a nanotube originate from opposite sides of the $K$ point.

Within each branch, the Raman intensity depends systematically on the chiral angle. It is small for  tubes close to the armchair direction (inner position) and first increases with decreasing chiral angle. Close to the zig-zag direction (outer positions) the intensity is small again. This is explained by the chiral-angle dependence of both the electron-phonon coupling and  the strength of the optical transitions.\cite{telg04,reich05b} The electron-phonon coupling is stronger for zig-zag tubes than for armchair tubes, explaining the weaker signal for the close-to-armchair tubes.\cite{machon05} The luminescence, on the other hand, was observed to decrease for close-to-zig-zag tubes\cite{bachilo02}, in particular for the $\nu=+1$ branches.\cite{reich05b} Although we do not know directly the chirality dependence of absorption strength into the second semiconducting transitions $E_{22}^S$, we can assume that it has an opposite dependence on chiral angle than the electron-phonon coupling, explaining the decrease of the resonant RBM signal towards the zig-zag direction.

In the last column of Tables~\ref{tab_alles_sem} and \ref{tab_alles_met} we show the intensity of  the measured RBM signal normalized to the Raman signal of CaF$_2$ and BaF$_2$ and divided by the Bose-Einstein occupation number. These values are proportional to the Raman susceptibility.

\section{Dependence on the type of surfactant}\label{sec_sds_sdbs}

In the previous sections we assumed the RBM frequency and the optical transition energies to reflect only the intrinsic properties of carbon nanotubes. Now we address the dependence of their properties on the surfactant (SDS or SDBS), \emph{i.e.}, the environment of the tube. The surfactant has a small influence on the position of the experimental data points in the Kataura plot both along the frequency (diameter) and the excitation energy axes. Besides the fundamental interest in the interaction between a nanotube and its surrounding, environment-related effects can affect a nanotube assignment based on a \emph{single} Raman spectrum and the RBM frequency alone (see Sect.~\ref{sec_how_to}). The pattern-based assignment, however, is unaffected by the shifts, because they are small and do not fundamentally change the experimental patterns.

\begin{figure}[t]
\resizebox{8cm}{!}{
\includegraphics*{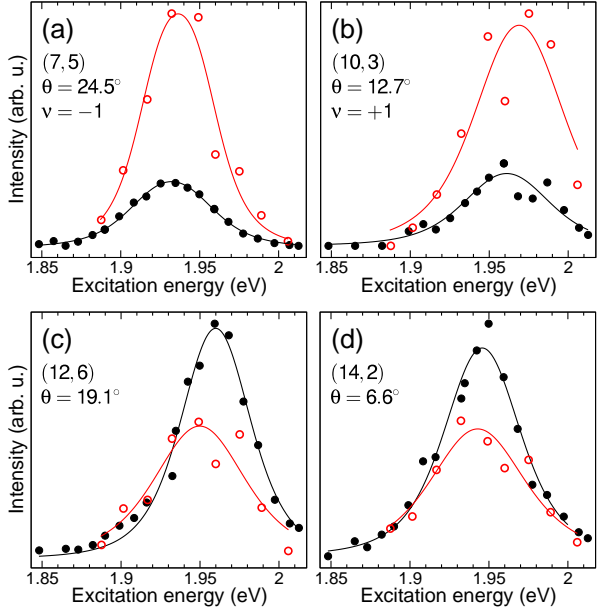}
}
\caption{(Color online) Resonance profiles of the RBM   for
  surfactants SDS (open red dots) and SDBS (filled black dots).  The chiral indices, the chiral angle $\theta$, and  the
   family $\nu$ are given. Semiconducting tubes, (a) and (b), show a small upshift in the
  position of resonance maximum for SDS. In metallic tubes, (c) and (d), the shift of the transition energies is in opposite direction.  The RBM intensity of
  metallic tubes is stronger in the SDBS sample, while the intensity of 
  semiconducting tubes is stronger in the SDS sample. Since the concentration of nanotubes in both solutions is not known exactly, we cannot quantify the absolute intensities. Comparing the relative intensities between the two surfactants,  they change in opposite directions for metallic and semiconducting tubes. }
\label{resProfSDSSDBS}
\end{figure}

To analyze the surfactant-induced changes in the Raman spectra, we recorded resonance profiles
for both surfactants with excitation energies between 1.85\,eV and 2.2\,eV,  see Fig.~\ref{resProfSDSSDBS}. In this region, the laser energies are in resonance with both metallic and semiconducting  tubes. 
We observe changes of $(i)$ the transition energies, $(ii)$ the Raman intensities, and $(iii)$ the RBM frequencies.

For semiconducting tubes [Fig.~\ref{resProfSDSSDBS}\,(a) and (b)], the transition energies shift by $\approx 5-10$\,meV to larger values in the SDS sample (open dots). In addition, 
the RBM signal of semiconducting tubes is stronger in the SDS sample than in the SDBS sample (closed dots). 
The behavior of  metallic tubes is opposite: their RBM intensity is larger in the SDBS sample. The shift of transition energies is, if detectable, of similar magnitude  as in semiconducting tubes but in opposite direction, \emph{i.e.}, the  transition energies of metallic tubes are slightly larger with SDBS as surfactant, see also Table~\ref{ResFitComp}. These changes are, however, small, and in particular for several tubes of  the metallic (15,0) branch within the experimental error.
 
\begin{figure}[t]
\begin{center}
\resizebox{8cm}{!}{
\includegraphics*{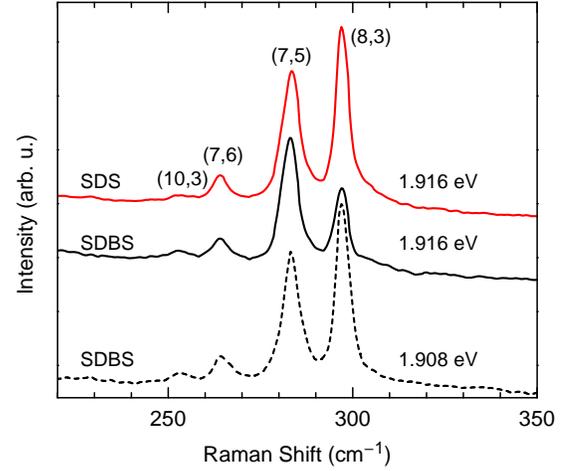}
}
\caption{(Color online) RBM spectra of nanotubes dispersed in D$_2$O using SDBS (black) and SDS
    (red; gray) as surfactants at excitation energy  1.916\,eV (solid lines) and 1.908\,eV (dashed lines). The spectra are normalized to the RBM amplitude of the (7,5) tube. 
}
  \label{SDSuSDBSat647}
\end{center}
\end{figure}

\begin{figure}[t]
\resizebox{8cm}{!}{
  \includegraphics*{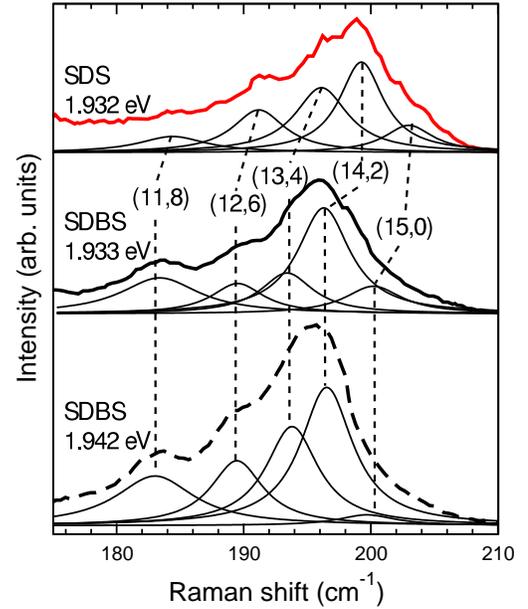}
}
 \caption{(Color online) Metallic part of the RBM spectrum. Red: SDS sample;
  black: SDBS sample. Thin lines show the fit of the RBM peaks by Lorentzians. The peaks are shifted to higher frequencies in the SDS sample.
}
  \label{peakshiftmet}
\end{figure}

In Fig.~\ref{SDSuSDBSat647} and \ref{peakshiftmet} we compare the RBM spectra for both surfactants at the same laser energy in the region of semiconducting and metallic tubes, respectively. The RBM frequency of semiconducting tubes is the same in both surfactants within experimental error (red and black curves in Fig.~\ref{SDSuSDBSat647}). The relative RBM intensities, in particular of the (8,3) and the (7,5) tube, are different in these spectra, reflecting the small shift of transition energy and hence of the resonance condition. The original intensity ratio in the SDS sample at $E_l=1.916$\,eV is recovered if the SDBS sample is excited at a slightly lower energy ($E_l=1.908$\,eV, dashed curve).  This shift in laser energy compensates for the shift in the optical transition energy of semiconducting tubes.
In metallic tubes we observe an upshift of the RBM frequencies in the SDS sample by about 2\,cm$^{-1}$, see Fig.~\ref{peakshiftmet}. 

The dependence of the RBM frequncy and intensity in metallic and semiconducting tubes on the type of surfactant agrees with the observation by Strano \emph{et al.}\cite{strano03e} of selective functionalization of metallic tubes. They found a decrease of the absorption strength for the metallic $E_{11}^M$ transitions, resulting from functionalization with tetrafluoroborate salt and formation of covalent bonds. Simultaneously, the RBM  shifted to larger frequency. We can thus interpret our results as due to an interaction between the surfactant and the nanotube, which is stronger for SDS than for SDBS. Although it is unlikely that a  covalent bond forms as in the case of Ref.~\onlinecite{strano03e},  an electron transfer from the metallic tubes to the surfactant might occur. The Raman intensity decreases in SDS as the resonant absorption becomes weaker, simultaneously the interaction leads to a larger RBM frequency. From our data we cannot detect such a difference in the interaction for semiconducting tubes, as the RBM is constant when changing the surfactant. 

\begin{figure}[t]
\resizebox{8.5cm}{!}{
\includegraphics*{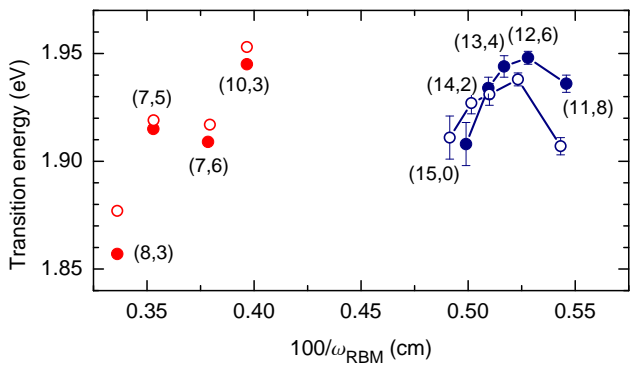}
}  
  \caption{(Color online) Section of the Kataura plot showing the transition energy
    \emph{versus} inverse RBM frequency of nanotubes  in
    two different surfactants, SDS (open dots) and SDBS (closed 
    dots). Semiconducting tubes (gray, red) show a uniform shift of the
    transition energies. Metallic tubes (dark, blue) are
    shifted in energy and RBM frequency. 
}
  \label{branchSDSvSDBS}
\end{figure}

The small shift in transition energies might be to first approximation assigned to a change in the dielectric environment. It is for several tubes within the range of experimental error. We have no explanation yet for the shift in opposite direction for metallic and semiconducting tubes.

Izard \emph{et al.}\cite{izard05} studied the development of RBM spectra from bundled tubes to bundles wrapped by SDS and to individual tubes in SDS. They also observed an upshift of the RBM due to wrapping by SDS which they assigned to pressure induced by the surfactant.  The metallic tubes appear to be more sensitive to the surfactant, as the RBM shift is in general larger than for semiconducting tubes, in agreement with our results. Izard \emph{et al.}\cite{izard05} observed changes in the relative RBM intensities as well, which they ascribed to a selective exfoliation process. From our data, we rather suggest a small change in resonance condition.

\begin{table}[tb]
\begin{ruledtabular}
\begin{tabular}{ccccccc}
     &\multicolumn{2}{c}{SDBS}   &\multicolumn{2}{c}{SDS}&  &  \\ 
tube   & $\omega_{\mathrm{RBM}}$ & $E_{ii}$ & $\omega_{\mathrm{RBM}}$ & $E_{ii}$ & $\Delta \omega$ & $\Delta E$ \\
   &    (cm$^{-1}$) & (eV) & (cm$^{-1}$) & (eV) & (cm$^{-1}$) & (meV) \\ \hline
   \multicolumn{6}{l}{metallic nanotubes}\\\hline
(15,0)       & 200         & 1.91    & 203        & 1.91    & $-$3           & $0\pm 20$ \\
(14,2)       & 196         & 1.93    & 199         & 1.93    &  $-$3          & $0\pm 10$ \\ 
(13,4)       & 193          & 1.94    &   196       & 1.93    &   $-$3         & 10$\pm 10$ \\ 
(12,6)       & 189.4      & 1.948    & 191.1       & 1.938    & $-$1.7           & 10$\pm 6$\\ 
(11,8)      & 183.2     & 1.936    & 184.1        & 1.906    & $-$0.9           & 30$\pm 8$ \\ \hline
\multicolumn{6}{l}{semiconducting nanotubes}\\\hline
(8,3)       & 297.5     & 1.857   & 297.5         & 1.877   & 0.0            & $-$20$\pm 13$ \\ 
(7,5)       & 283.3      & 1.915   & 283.2          & 1.919   & 0.1            & $-$4$\pm 4$ \\
(7,6)       & 264.2      & 1.909   & 263.6           & 1.917   & 0.6            & $-$8$\pm 6$ \\ 
(10,3)      & 252.1        & 1.945   & 252.1           & 1.953   & 0.0            & $-$8$\pm 6$ 
\end{tabular}
\end{ruledtabular}
\caption{
Comparison of transition energies and RBM frequencies for  different surfactants (SDS and SDBS).  All
transition energies are obtained from resonance profiles. The experimental errors for the first three tubes of the (15,0) branch are larger than for the majority of our data.
}
\label{ResFitComp}
\end{table}

Our assignment of Sect.~\ref{sec_assign} is not  
affected by the surfactant-induced variation in the transition energies and RBM frequencies.
Figure~\ref{branchSDSvSDBS}  shows a small section of the Kataura plot with data from SDS 
(open dots) and from SDBS (closed dots). The differences in excitation energies and RBM frequencies are minor on the scale of the Kataura plot. The
most important criteria for the assignment are the RBM frequency patterns and the
number of tubes  within a branch. Figure~\ref{branchSDSvSDBS} shows that the small variations of the RBM do not change these systematics. In particular, the changes are too small to shift the data to a different branch. Therefore, the assignment is valid for both types of  surfactant.

\section{How to assign $(n_1,n_2)$ in a Raman experiment}\label{sec_how_to}

A great need for nanotube research is to identify the chirality of a tube before performing an experiment. Ideally, the method is non-destructive, does not require special equipment or substrates, works for semiconducting and metallic tubes as well as individual tubes and bulk samples (nanotubes in solution or bundled tubes). It should also give reliable results regardless of the tube environment. We discussed possible ways for identifying carbon nanotubes in the introduction of this paper. Raman scattering meets many of the requirements for becoming one of the prime assignment methods for single-walled carbon nanotubes.

Most Raman-based assignments of individual and bundled tubes relied mainly on the $\omega_{\mathrm{RBM}}\propto 1/d$ relationship using one value for $c_1$ and $c_2$ or the other, see the review in Ref.~\onlinecite{reich04}. In this paper we showed that the RBM frequency \emph{alone}  will never be sufficient for assigning the chirality, because it depends on the environment of the tubes. Although the changes in the RBM frequencies are small between different surfactants (Sect.~\ref{sec_sds_sdbs}) and also between bundled and surfactant-coated tubes,\cite{fantini04,oconnell04,izard05} they are large enough to change an assignment that uses the RBM frequencies as the only input, see also Ref.~\onlinecite{epaps}. Therefore, a Raman-based assignment of an individual tube, suspended or on a substrate, nanotubes in solution, nanotube bundles and so forth should always use the combined information of RBM frequency and excitation energy.

Once the full resonance Raman experiment has been performed and the assignment of $\omega_{\mathrm{RBM}}$ to the chiral index has been found (Sec.~\ref{sec_assign}), the chiral indices  of a sample containing tubes with similar diameters can be  determined from one single Raman spectrum. Environment-related effects often can be taken into account by estimating the changes in the transition energies. In this section we explain the procedure and show how the RBM peaks in a given Raman spectrum can be assigned to $(n_1,n_2)$.

\begin{figure} 
\begin{center}
\resizebox{8cm}{!}{
\includegraphics*{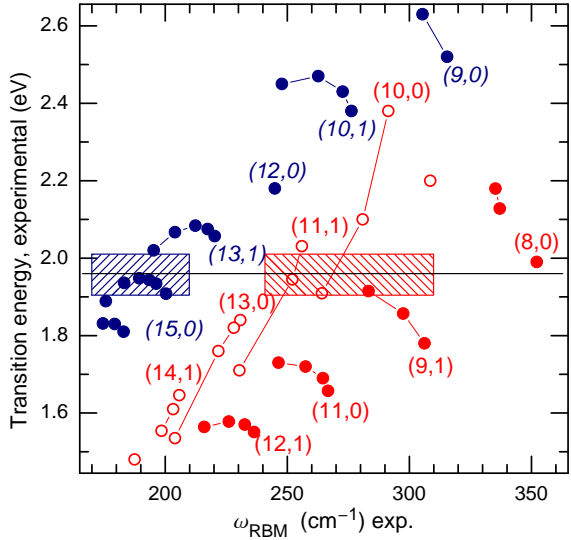}
}
 \end{center}
\caption{(Color online) Experimental Kataura plot: transition energies \emph{vs.} RBM frequencies, from Tables~\ref{tab_alles_sem} and \ref{tab_alles_met}. The shaded regions correspond to the RBM peaks in Fig.~\ref{spektrum_assign}.}
\label{kataura_exp}       
\end{figure}

\begin{figure}[t] 
\begin{center}
\resizebox{8cm}{!}{
\includegraphics*{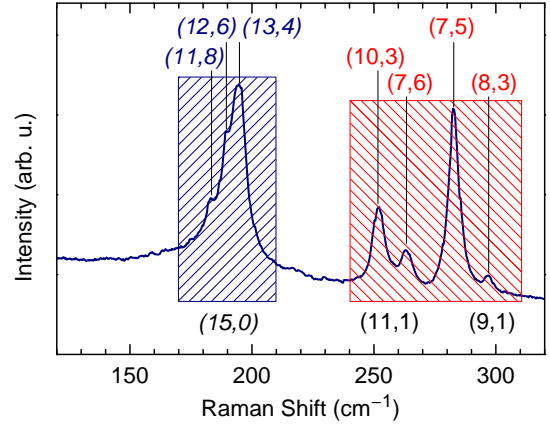}
}
 \end{center}
\caption{(Color online) RBM spectrum at $E_l=1.96$\,eV. The shaded regions indicate groups of metallic and semiconducting tubes; the chiral indices at the bottom give the first element of the corresponding \emph{laola} in the Kataura plot (Fig.~\ref{kataura_exp}). The chiral indices at the top show the assignment of the strongest peaks.}
\label{spektrum_assign}       
\end{figure}

For simplicity, we plot the experimental data of Fig.~\ref{kataura_1} (Tables~\ref{tab_alles_sem} and \ref{tab_alles_met}) now with the RBM frequency along the $x$ axis, see Fig.~\ref{kataura_exp}. The procedure of the assignment is as follows:
\begin{enumerate}
\item
Record a Raman spectrum at a given excitation energy $E_l$ and determine the RBM frequencies $\omega_{\mathrm{RBM}}$ (Fig.~\ref{spektrum_assign}).
\item
Identify  groups of $3-5$ close-by RBM peaks as indicated in Fig.~\ref{spektrum_assign} by the shaded areas. The members of these groups form a branch (\emph{laola}) in the Kataura plot.
\item
Find the excitation energy $E_l$ in the experimental Kataura plot (horizontal line at $1.96$\,eV in Fig.~\ref{kataura_exp}).
\item
Find along the $x$-axis  of the experimental Kataura plot the regions of observed RBM frequencies (shaded areas). The branches closest to $E_l$ within these regions most likely contribute to the RBM spectrum. 
\item
Compare all RBM frequencies in detail with the experimental Kataura plot and the sequence of tubes in a branch to find the final assignment, see Fig.~\ref{spektrum_assign}. Compare the number of tubes within this branch to  how many of them are observed.
\end{enumerate}

In the example given in Figs.~\ref{kataura_exp} and \ref{spektrum_assign}, we identify the left and the right RBM groups as metallic and semiconducting tubes, respectively. Since the width of the resonance profiles is typically around $60$\,meV, we assume that we can observe tubes from a window of approximately $100-200$\,meV width around the excitation energy. These areas are indicated in Fig.~\ref{kataura_exp}. For the metallic tubes, mainly the (15,0) branch contributes to the spectra. In the region of semiconducting tubes, the members of several branches are close to the excitation energy.  We identify the peaks as resulting from tubes of the (11,1), (10,0), and (9,1) branches. In the final step, we assign the strongest peaks to the (11,8,), (12,6), and (13,4) tube (metallic region) and to the (10,3), (7,6), (7,5), and (8,3) tube (semiconducting region), as indicated in Fig.~\ref{spektrum_assign}. The remaining tubes of these branches, such as the (15,0) tube, are weaker shoulders of the strongest peaks and become evident when changing the laser wavelength. Others, like the members of the (10,0) branch, are outside the resonance window. Thus not all members of each branch are expected in the same single Raman spectrum; because of the chirality-dependence of the Raman cross section some tubes might not be observed.

The most relevant piece of information is already obtained in the fourth step, \emph{i.e.}, by identifying the correct branches. Many properties are similar for  tubes of  the same branch. In contrast, they differ strongly for tubes of different branches even if the diameters or transition energies are almost the same. For example, the (7,5) and the (7,6) tube have very similar $E_{ii}$, but the (7,6) tube gives a much weaker RBM signal. This is due to the different strength of the  electron-phonon  coupling for $\nu=-1$ and $\nu=+1$ tubes\cite{machon05}, as explained in Sec.~\ref{rbm_intensities}.

In Raman measurements on a single, individual nanotube, the chiral index  is found by the same method. For this type of samples the difficulty is to obtain an observable RBM signal, because of the narrow resonance window of the RBM. One, two or even three laser energies might be needed to find the resonance window of a particular tube. The \emph{laola} groups---typical for samples with different chiralities---that allowed to identify branches, are absent in individual tubes. Nevertheless, following the procedure described above, the choice of possible chiralities can be narrowed to one or two tubes.
If the ambiguity concerns two nanotubes from the same branch, a further refinement of the assignment is not necessary.  Two neighboring tubes in the same branch are too similar in properties to easily distinguish between them. In turn, this implies that neither fundamental studies nor applications benefit much from narrowing down the choice. If, on the other hand, a single Raman spectrum is insufficient to distinguish between two branches, the ambiguity arose between two tubes of different family---metallic, $\nu=+1$ or $\nu-1$ semiconducting. In this case the assignment should be verified by using a different excitation energy or by combining Raman scattering with a second assignment technique. 

\begin{figure}[t] 
\begin{center}
\resizebox{8cm}{!}{
\includegraphics*{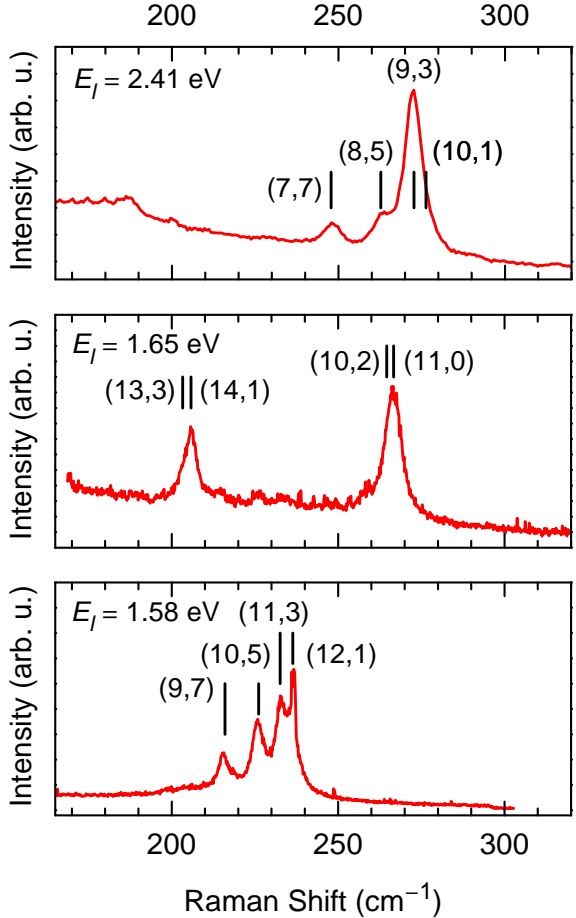}
}
 \end{center}
\caption{RBM spectra with chiral-index assignment at several standard laser lines, 514\,nm, 752\,nm, and 785\,nm. The peaks in the middel panel consist of two RBMs each which can only be resolved when changing the excitation energy. For the spectrum at 633\,nm (1.96\,eV) see Fig.~\ref{spektrum_assign}. }
\label{beispiele}       
\end{figure}

In Sec.~\ref{sec_sds_sdbs} we showed that changing the environment of the nanotubes did not affect the RBM frequencies of semiconducting tubes and only slightly those of metallic tubes. This is consistent with the observation of only small RBM changes in bundles \emph{vs.} separated tubes and in tubes in several different surfactants.\cite{fantini04,oconnell04,izard05} On the other hand, the transition energies $E_{ii}$ appear to be more sensitive to the nanotube environment.\cite{fantini04,oconnell04,izard05,moore03} Therefore, the empirical values of $E_{ii}$ given in Tables~\ref{tab_alles_sem} and \ref{tab_alles_met} are strictly valid only for nanotubes in SDS/SDBS and should be used with care for other types of samples. Ideally, one would perform the full resonance Raman experiment once for each tube environment, e.g., different surfactants, bundled tubes, individual tubes on a substrate, individual suspended tubes in air, etc. Some of these data have been reported in the literature, see for instance Refs.~\onlinecite{fantini04,oconnell04}. In addition, photoluminescence data can be used for $E_{ii}$ in different samples. The Raman-based assignment procedure, however, is always the same as described in this section. In particular, if entire branches are observed, the data presented in Tables~\ref{tab_alles_sem} and \ref{tab_alles_met} can be readily used, taking into account changes in the $E_{ii}$. As the RBM frequencies vary only slightly, the Raman-based assignment is much more stable against changes in the nanotube environment than an assignment based on the $E_{ii}$ alone.

In standard Raman setups often just a few laser lines are available. To facilitate an assignment, we show in Fig.~\ref{beispiele} the RBM spectra for the most common laser lines (514\,nm, 752\,nm, and 785\,nm) together with the chiral indices. These spectra can be diretly compared to Raman spectra taken on HiPCO tubes in solution with standard equipment and used for a simple assignment. Two tubes with very similar diameters are sometimes difficult to resolve from a single Raman spectrum, see the 752\,nm spectrum in Fig.~\ref{beispiele}. Apparently only two tubes contribute. From the excitation-energy dependent measurements we know that each peak in the middle pannel of Fig.~\ref{beispiele} is, in fact, composed of two RBM lines.

\section{Summary}

In summary, we  presented a chiral-index assignment for carbon nanotubes from  resonant Raman scattering. The assignment is independent of any additional parameters, but it is based on pattern recognition. 
The two pieces of information that are required for this assignment are the frequency of the radial breathing mode and the energy of an optical transition (here $E_{22}^S$ and $E_{11}^M$). They constitute an experimental Kataura plot where all chiral indices are systematically grouped into so-called branches with neighboring indices given by $(n_1^\prime,n_2^\prime)=(n_1-1,n_2+2)$. Because of these systematics, the  assignment remains the same even if parameters in the calculation of $E_{ii}$ or the diameter change or if the experimental values vary due to slightly different experimental conditions.  $(n_1,n_2)$ is assigned to \emph{experimental} RBM frequencies and transition energies, irrespective  of changes in the theoretical description. We consider all measured transition energies to be excitonic energies, as excitonic effects dominate the optical spectra in carbon nanotubes. 

We derived the parameters $c_1=215\pm2$\,cm$^{-1}$nm and $c_2=18\pm2$\,cm$^{-1}$ from our assignment for the RBM-diameter relation. These values vary depending on the type of sample and on the details of the diameter calculation. The RBM intensities are in general stronger for $\nu=-1$ nanotubes than for $\nu=+1$ tubes for the $E_{22}^S$ transitions. They decrease from their maximum around $\theta\approx 10-15^\circ$ towards both the armchair and the zig-zag direction. These results are in good agreement with \emph{ab-initio} calculations of the electron-phonon coupling.\cite{machon05} The intensities also depend on the type of surfactant in our samples with different behavior for metallic and semiconducting tubes. For metallic tubes, we observed a stronger interaction with SDS, and an upshift of the RBM frequencies.

Finally, we provided a description on how to find a chiral-index assignment from a single Raman spectrum. For samples with similar tube diameters in a similar environment, the experimental and empirical data given in Tables~\ref{tab_alles_sem} and \ref{tab_alles_met}, and in Ref.~\onlinecite{epaps} can be used for a straightforward assignment. Changes in the tube environment usually affect mainly the optical transition energies, which can be taken into account for an assignment.
We stress that the RBM frequencies \emph{alone} are insufficient for an assignment. It  should always be based on the combined information of RBM frequency and excitation energy. Taking into account these two pieces of information results in a robust and reliable assignment based on Raman spectroscopy.

\section*{Acknowledgment}
We thank F. Hennrich for providing us with the  samples.
%This work was supported by the DFG under grant number Th 662/8-2.
S. R. was supported by the Oppenheimer Fund and  Newnham College.

\end{document}